\documentclass[aps,prl,showpacs,twocolumn,amssymb]{revtex4}
\usepackage{graphicx}
\begin{document}

\draft

\title{Transverse conductivity in the sliding CDW state of NbSe$_3$ }

\author{A.~A.~Sinchenko},
\address{National research nuclear university (MEPhI), 115409 Moscow,
Russia}
\author{P.~Monceau},
\address{Institut NEEL, CNRS and Universit\'{e} Joseph Fourier, BP 166,
38042 Grenoble, France}
\author{T.~Crozes}

\address{Institut NEEL, CNRS and Universit\'{e} Joseph Fourier, BP 166,
38042 Grenoble, France}

\date{\today}

\begin{abstract}

The dynamical properties of longitudinal and transverse conduction
of NbSe$_3$ single-crystals have been simultaneously studied when
the current is applied along the $b$ axis (chain direction). In the
vicinity of the threshold electric field for CDW sliding, the
transverse conduction sharply decreases. When a rf field is
applied, voltage Shapiro steps for longitudinal transport are
observed as usual, but also current Shapiro steps in the transverse
direction. The possible mechanisms of this effect are discussed.

\end{abstract}

\maketitle

One of the most interesting property of quasi-one
dimensional-conductors with a charge-density-wave ground state (CDW)
is their nonlinear electronic transport associated with the
collective motion of the CDW above a depinning threshold electric
field, $E_t$ \cite{Gruner}. In Refs.\cite{pokrovskii07} it
was shown that when a voltage near threshold is applied to a crystal
of o-TaS$_3$ which is free to distort, the crystal twists by a small
rotation angle achieving several degrees as a result of torsional
strain. This effect was associated with surface shear of the CDW
coupled to the crystal shear that obviously results from some
deformation of the CDW in the transverse direction. These data
indicate that some peculiarities in transverse properties take place
when the longitudinal electric field overcomes $E_t$.

In the present letter we report for the first time observation of
dynamical effects in transverse conduction in NbSe$_3$ when the
CDW is sliding along the chains.

NbSe3 is a layered quasi-one-dimensional (Q1D) conductor exhibiting
two incommensurate charge-density wave (CDW) transitions at
$T_{p1}=145$ K and $T_{p2}=59$ K \cite{Gruner}. The Peierls
transitions in this material are not complete and ungapped carriers
remain in small pockets at the Fermi level. As a result, NbSe$_3$
keeps metallic properties down to the lowest temperatures. The
crystal lattice of NbSe$_3$ is monoclinic with the $b$ axis being
parallel to the CDW chains and corresponding to maximum
conductivity. NbSe$_3$ single crystals have a ribbon shape with a
long size coinciding with the CDW chain direction and the width
along the c-axis.

\begin{figure}[t]
\includegraphics[width=8cm]{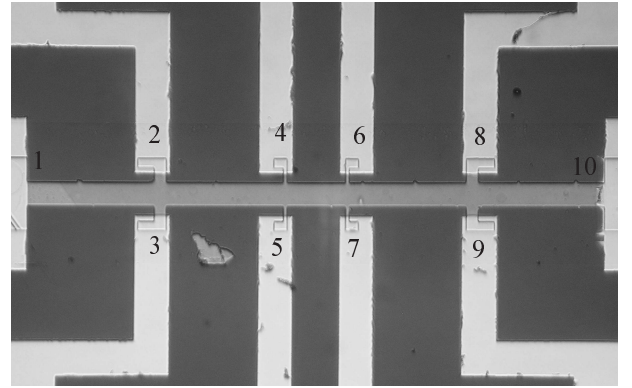}
\caption{\label{F1}(color online) Image of one of the sample.}
\end{figure}

For the experiment we selected only high quality NbSe$_3$
single-crystals with a thickness (0.3-0.4) $\mu$m and a width 60-100
$\mu$m. The crystals selected were cleaned in oxygen plasma, glued
on sapphire substrates and patterned into a ten-probe configuration
from the single crystal itself with the help of electron
lithography. A SEM image of one of such a structure is shown in
Fig.\ref{F1}. The width of the sample is 20 $\mu$m. The distance
between probes 1-2; 2-4; 4-6; 6-8; 8-10 are 100; 100; 50; 100 and
100 $\mu$m correspondingly. The width of contacts 2,3,8 and 9 are 10
$\mu$m; width of contacts 4,5,6 and 7 are 3 $\mu$m. Three such
structures were prepared and measured.

In the Hall-bar geometry (Fig.1) without any magnetic field, in the
ohmic regime, with the current applied along the chain axis
\cite{Sinch11}, we have measured a finite voltage,$V_{tr}$, across
the potential leads. In the normal state between room temperature
and $T_{p1}$, this voltage is weak;  its magnitude and sign  may
depend on the exact current direction with a temperature dependence
slightly positive or negative (even nul), as it was checked by
injecting the transport current through the two pairs of contact,
2-3 and 8-9, each pair being connected through variable resistors.
$V_{tr}$ is also dependent of the sample inhomogeneity. Then, for
studying properties of the transverse conduction in the longitudinal
CDW sliding state, it is appropriate to have well defined transverse
components of the electronic transport. With this respect, we used
the pairs of contacts 2-9 or 3-8 as  current electrodes.  For
transverse voltage, $V_{tr}$, measurements were performed through
the opposite potential probes 4-5 and 6-7 . The longitudinal drop of
voltage, $V_L$, was measured on 4-6 or 5-7 contacts. The disturbing
voltage signals (thermoelectric etc.) were eliminated by reversing
the transport current in the measurements of the temperature
dependencies $V_L(T)$ and $V_{tr}(T)$. Although we measure a
transverse voltage, it is more convenient to use resistance rather
than voltage in the analysis of the data; then we define
$R_L=V_L(T)/I$ and $R_{tr}=V_{tr}(T)/I$.  For studying nonstationary
effects a radiofrequency (rf) current was superposed on the dc
current using contacts 1 and 10 via two capacitors.

The temperature dependencies $R_L(T)$ and $R_{tr}(T)$ at different
currents are shown in Fig.2. The qualitatively same dependencies
were observed for the three structures we measured. In the ohmic
regime, the $R_L(T)$ dependence at $I=0.02$ mA (inset in Fig.2 (b))
has the conventional shape for the resistance variation along the
conducting chains in NbSe3 in the static pinned state. As can be
seen, a qualitatively different dependence takes place for
$R_{tr}(T)$ at the same current (inset in Fig.2 (b)). The
temperature dependence of $R_{tr}$ does not follow at all the
resistance $R_c(T)$ measured at constant current applied along the
$c$-axis direction \cite{OngBrill} but exhibits a sharp drop at
$T_{p1}$ and $T_{p2}$. That indicates moreover a negligible
contribution to $V_{tr}$ from the longitudinal voltage variation,
which may have resulted possibly due to a misalignment of potential
probes or to current injection not strictly parallel to the
conducting chains.

The appearance of this electric potential normal to the transport
current has been attributed to fluctuations of the critical
temperature of the Peierls transition. It has also been demonstrated
that the transverse voltage is proportional to the derivative of the
longitudinal voltage, $V_{tr}\propto dR_L/dT$ (Fig.4 in
Ref.\cite{Sinch11}).

\begin{figure}[t]
\includegraphics[width=8cm]{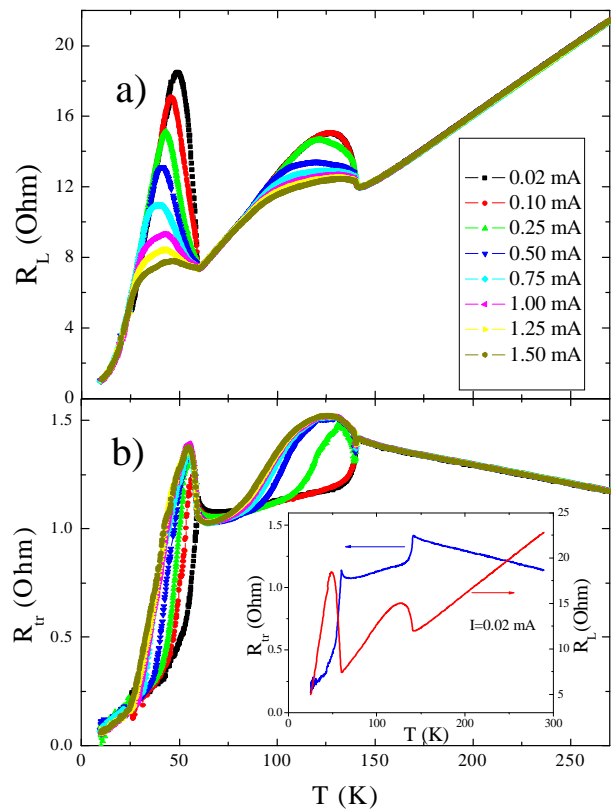}
\caption{\label{F2}(color online) Temperature dependencies of a)
longitudinal resistance, $R_b=V_L/I$, for sample \#1, at different
currents; b) transverse resistance $R_{tr}=V_{tr}/I$ at the same
values of current. Inset in fig.2 (b) shows $R(T)$ for longitudinal
(red curve) and transverse directions (blue curve) in the static CDW
state.}
\end{figure}

CDWs start to slide for transport current $I\gtrsim40$ $\mu$A
inducing significant changes in both transverse and longitudinal
resistances. As can be seen in Fig.2 (a),  maxima of $R_L$ observed
below $T_{p1}$ and $T_{p2}$ decrease as usual, indicating the
increase of conductivity due to the CDW sliding. The opposite
picture is observed in the $R_{tr}$ behaviour (Fig.\ref{F2} (b)): a
strong increase of the transverse voltage takes place below $T_{p1}$
and $T_{p2}$ indicating the decrease of conductivity. As the CDW
cannot slide  transversely to chains \cite{Ayari}, the electronic
transport in this direction can only be attributed to uncondensed
carriers and/or to the conversion CDW - normal carriers.

\begin{figure}[t]
\includegraphics[width=8cm]{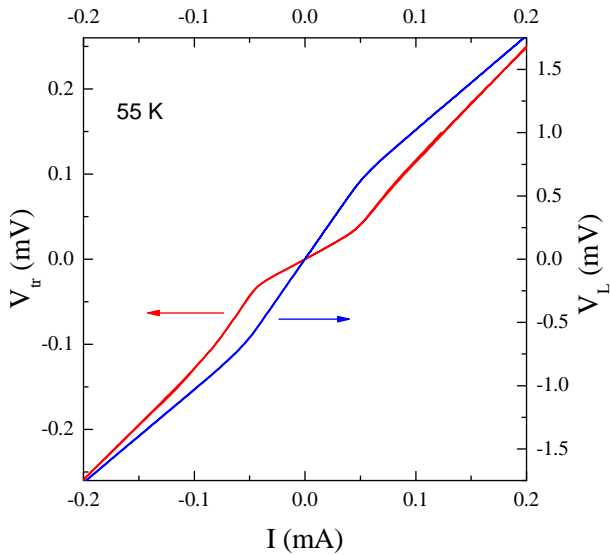}
\caption{\label{F3}(color online) $V{tr}(I)$ (red curve) and
$V_L(I)$ (blue curve) measured at $T=55$ K for sample \#3.}
\end{figure}

\begin{figure}[t]
\includegraphics[width=8cm]{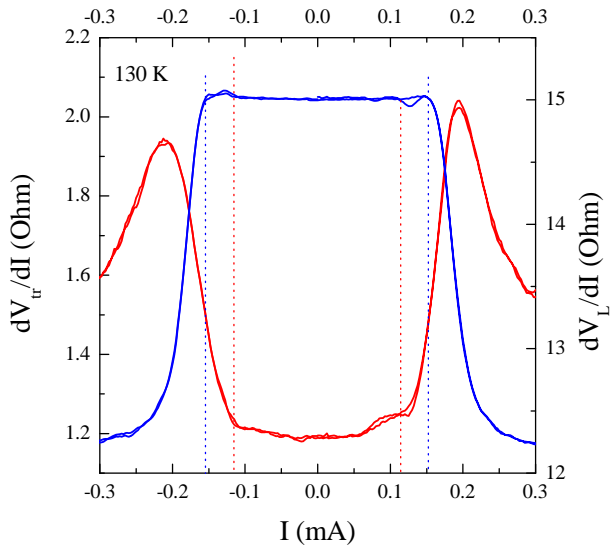}
\caption{\label{F4}(color online) $V_{tr}/dI(I)$ (red curve) and
$dV_L/dI(I)$ (blue curve) measured at $T=130$ K for sample \#3.
Dotted lines indicate the beginning of increase of the transverse
voltage and the decrease of longitudinal voltage.}
\end{figure}

In Fig.\ref{F3}, we have drawn the current-voltage characteristics
(IVc) in the transverse $V_{tr}(I)$ (red curve) and the longitudinal
$V_L(I)$ (blue curve) directions measured simultaneously for the low
temperature CDW ($T=55$ K). Fig.\ref{F4} shows the derivatives
$dV_{tr}/dI(I)$ (red curve) and $dV_L/dI(I)$ (blue curve) for the
high temperature CDW ($T=130$ K). The curves for the longitudinal
direction are typical IVc: the ohmic behavior is observed at low
current and the excess CDW current appears above the threshold current
$I_t$ resulting from CDW sliding. For the transverse direction the
IVc present the following features:

i) For both CDWs in the vicinity of the threshold current, $I_t$,
defined as the beginning of the decrease of $dV_L/dI$, a sharp
increase of $dV_{tr}/dI$ is observed. Below this current IVc have a
ohmic character.

ii) The increase of $dV_{tr}/dI$ (indicated by red dotted lines)
occurs at a current which is less than the threshold current for CDW
sliding (indicated by blue dotted lines).

\begin{figure}[b]
\includegraphics[width=8cm]{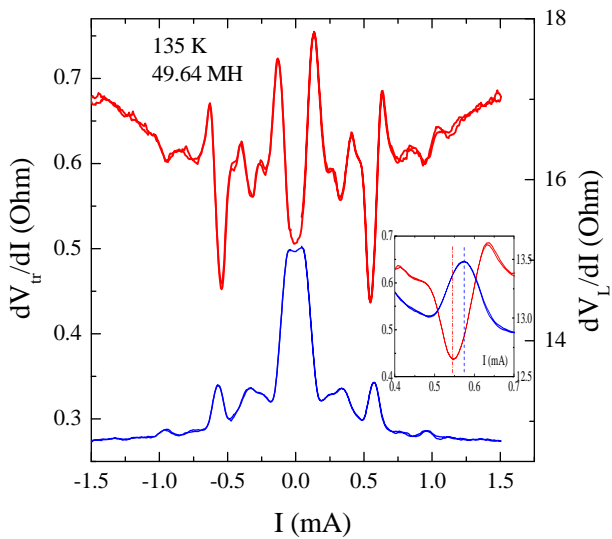}
\caption{\label{F5}(color online) $dV_{tr}/dI(I)$ (red curve) and
$dV_L/dI(I)$ (blue curve) measured at $T=135$ K for sample \#3 under
application of a rf field with frequency 49.64 MHz. Inset shows the
exact position of the corresponding transverse (red) and
longitudinal (blue) Shapiro steps}
\end{figure}

The joint application of dc and rf driving fields leads to
appearance of harmonic and subharmonic Shapiro steps in the
longitudinal dc IV characteristics \cite{Gruner}. In the present
work we have observed Shapiro steps in both longitudinal and
transverse directions below $T_{p1}$ and $T_{p2}$. Fig.\ref{F5}
shows $dV_{tr}/dI(I)$ (red curve) and $dV_L/dI(I)$ (blue curve) at
135 K under application of a rf field with a frequency 49.64 MHz. As
can be seen, Shapiro steps for longitudinal transport appear in the
differential resistance $dV_L/dI(I)$ characteristic as spikes, as
usual;  that corresponds to voltage steps. On the contrary,  in the
$dV_{tr}/dI(I)$ characteristic, minima in the differential
resistance are observed, that corresponding to Shapiro current
steps. Note that, in our experimental conditions of a low rf power
and consequently without complete mode locking,   we observe Shapiro
steps  in transverse transport with a larger amplitude and  much
more pronounced features when compared to those in the longitudinal
direction.

Thus, data indicate that nonlinearity in $V_{tr}(I)$ occurs at a
smaller threshold field than in $V_L(I)$. It has to be noted that
the threshold field in elastic measurements, and particularly in the
shear compliance data, was also shown to be below that for changes
in transport measurements \cite{Staresinic01}. Voltage-induced
torsional strain is also initiated at voltage below the transport
longitudinal threshold, suggesting that the torsional crystal strain
is caused by deformations of the CDW, rather than by the CDW
current. This torsional strain was attributed to CDW wavefronts
being twisted, even without applied voltage and due, for instance to
contacts or defects \cite{Brill10}.

One may consider the possibility that current density inhomogeneity
results from lateral current injection through electrodes 2-9 or
3-8, in the vicinity of which electric field is stronger. Then,
depinning may occur  near these electrodes at a lower threshold than
that between electrodes 4-5 or 6-7. However, we have measured
$dV_L/dI(I)$ characteristics with current injection from contacts
2-8 and from 1-10 and found a difference less than 2\%. In addition,
a strong current inhomogeneity which should account for the
difference of around 20\% between longitudinal and transverse
threshold should lead the threshold characteristic to get smeared.
In contrast, as seen in Fig. 4,  we observe very sharp $dV/dI(I)$
characteristics longitudinally and transversally with well defined
threshold fields. That allows us to  conclude that current density
inhomogeneity in the measuring part of the sample cannot be the
explanation of the effect we are reporting.

The fact that $E_{t1}$ in the transverse direction is less than the
longitudinal $E_t$, but close to it, means that CDW deformations are
necessary for CDW sliding. This CDW deformation may be considered as
a part of sliding does randomly distributed impurities, being a
precursor to sliding. One can also propose the statement in which it
is the change in transverse coherence which triggers the
longitudinal CDW motion.Thus, when a longitudinal electric field is
applied up to $E_{t1}<E_t$ , the CDW is deformed up to a certain
critical value, above which the transverse CDW coherence is lost.
This proposed scenario is in agreement with results of
\cite{Brazovskii00, DiCarlo93} where, from high-resolution x-ray
scattering measurements, a strong reduction of the transverse
correlation was observed when the CDW moves. Thus, due to the loss
of transverse coherence in motion, CDW velocities are distributed
along filaments in the $b$-axis.

Due to the presence of impurities in or between chains, a
configuration similar to weak links (tunnel contact) between
adjacent CDW filaments can be realized. In such a configuration,
significant phase shift appears between CDW chains. It was
theoretically shown by Artemenko and Volkov
\cite{Artemenko97,ArtemenkoVolkov84} that the transverse current
comprises a term proportional to the cosine of the CDW phase
difference between neighbouring chains. Let first consider the
simple case when the CDW slides with different velocities along two
chains coupled by a weak link; then the phases vary with time and
alternating tunneling current is generated transversely to the chain
direction with a frequency depending on the longitudinal electric
field. When an external alternating signal acts on the sample, a
resonance will be observed for a fixed $V_{tr}$ if the frequency of
the external field and that of the characteristic oscillations
coincide. As a result current Shapiro steps will appear in the
transverse IV characteristics. This effect shows some analogy with
the time-dependent Josephson effect in weakly linked
superconductors. However, in a macroscopic sample, the CDW phase
fluctuates because the presence of impurities or takes different
values in different domains  which may wash out the time-dependent
phenomena. However, experimentally, we have observed more pronounced
features of Shapiro steps in the transverse direction than in the
longitudinal one (Fig.\ref{F5}). That may imply a specific
synchronisation between oscillations from all the weak links in the
sample.

It may have another possible explanation in which the alternating
tunneling current takes its origin at the electrodes 4, 5 or 6, 7;
at these places the CDW is time dependent along the chain axis but
not time dependent in the electrode where the CDW remains pinned ,
realizing thus a (nearly perfect) junction. However it has to be
envisaged that this macroscopic junction (size of 2-3 $\mu$m) is
composed of many microscopic weak links with characteristic
dimension of the order of the amplitude CDW coherence length, which
is the frame of the theoretical model of Artemenko and Volkov
\cite{ArtemenkoVolkov84}.

In summary, we have measured for the first time the dynamical
properties which occur in the transverse direction to the conducting
chains in NbSe$_3$ single-crystals, when the CDW slides along the
chains. At an electric field less than the longitudinal threshold
one for CDW sliding a sharp decrease in transverse conductivity
takes place;  that may result from induced phase shifts between CDW
chains. Under the joint application of dc and rf driving fields
pronounced current Shapiro steps in transverse transport have been
observed. The results were tentatively explained in the frame of
Artemenko-Volkov theory \cite{ArtemenkoVolkov84,Artemenko97}. Note
that it would be worth to consider the effect of the transverse
voltage we have measured in other nonlinear transport CDW phenomena
with induced CDW deformations.

\acknowledgements

The authors are thankful S.A. Brazovskii and especially to S.N.
Artemenko for helpful discussions of the experimental results and J.
Marcus for help in the sample preparation. The work has been
supported by Russian State Fund for the Basic Research (No.
11-02-01379-à), and partially performed in the frame of the CNRS-RAS
Associated International Laboratory between CRTBT and IRE "Physical
properties of coherent electronic states in coherent matter".The
support of ANR-07-BLAN-0136 is also acknowledged.

\end{document}